# Observation of ballistic avalanche phenomena in nanoscale vertical InSe/BP heterostructures

Anyuan Gao[1], Jiawei Lai[2], Yaojia Wang[1], Zhen Zhu[3], Junwen Zeng[1], Geliang Yu[1], Naizhou Wang[4,5], Wenchao Chen[6], Tianjun Cao[1], Weida Hu[7], Dong Sun[2,8], Xianhui Chen[4,5], Feng Miao[1*], Yi Shi[1*] & Xiaomu Wang[1*]

[1]National Laboratory of Solid State Microstructures, School of Physics, School of Electronic Science and Engineering, Collaborative Innovation Centre of Advanced Microstructures, Nanjing University, Nanjing 210093, China.

[2]International Centre for Quantum Materials, School of Physics, Peking University, Beijing 100871, China.

[3]Materials Department, University of California, Santa Barbara, CA 93106-5050, U.S.A.

[4]Hefei National Laboratory for Physical Science at Microscale and Department of Physics, University of Science and Technology of China, Hefei, Anhui 230026, China.

[5]Key Laboratory of Strongly Coupled Quantum Matter Physics, University of Science and Technology of China, Hefei, Anhui 230026, China.

[6]College of Information Science and Electronic Engineering, ZJU-UIUC Institute, Zhejiang University, Haining 314400, China.

[7]State Key Laboratory of Infrared Physics, Shanghai Institute of Technical Physics, Chinese Academy of Sciences, Shanghai 200083, China.

[8]Collaborative Innovation Centre of Quantum Matter, Beijing 100871, China.

*Corresponding author: E-mail: xiaomu.wang@nju.edu.cn (X.W.); miao@nju.edu.cn (F.M.); yshi@nju.edu.cn (Y.S.))

**Impact ionization, which supports carrier multiplication, is promising for applications in single photon detection[1] and sharp threshold swing field effect devices[2]. However, initiating impact ionization of avalanche breakdown requires a high applied electric field in a long active region, hampering carrier-multiplication with high gain, low bias and superior noise performance[3, 4]. Here we report the observation of ballistic avalanche phenomena in sub-mean free path (MFP) scaled vertical InSe[5]/black phosphorus (BP)[6-9] heterostructures[10]. We use these heterojunctions to fabricate avalanche photodetectors (APD) and impact ionization transistors with sensitive mid-IR light detection (4 μm wavelength) and steep subthreshold swing (SS) (<0.25 mV/dec). The devices show a low avalanche threshold (<1 volt), low noise figure and distinctive density spectral shape. Our transport measurements suggest that the breakdown originates from a ballistic avalanche phenomenon, where the sub-MFP BP channel support the lattice impact-ionization by electrons and holes and the abrupt current amplification without scattering from the obstacles in a deterministic nature. Our results provide new strategies for the development of advanced photodetectors[1, 11, 12] via efficient carrier manipulation at the nanoscale.**

The upper panel of Fig. 1a shows the schematic of our heterostructure device with electrical connections. It consists of a thin γ-rhombohedral InSe/BP heterostructure connected to bottom and top metal electrodes on substrate. The detailed fabrication processes are presented in the Methods Section. During our measurements, we define the biased electrode connected to BP (~10 nm) as the drain and the grounded electrode connected to InSe (also ~10 nm) as the source. Considering that the lateral resistance of unstacked InSe is much smaller than that of the junction (Supplementary Section 1a), carriers also mainly transport vertically along the nanoscale InSe/BP channel. As a result, n-type InSe and p-type BP[13-15] form a vertical vdW heterojunction. The bottom panel of Fig. 1a schematically shows the lattice structure at the junction interface. Notably, we assembled the InSe/BP junction in a glove box, resulting in a nearly perfect interface. In Fig. 1b, the high-resolution transmission electron microscope image verifies that the atomic stack is clean without the presence of any contamination or amorphous oxide even after all the device fabrication processes.

We first characterized the transport properties of our heterostructure devices. With proper gate voltage[16] (10 V here, to ensure a necessary low doping level of BP and InSe, see below for details), the vertical vdW junction presents a standard rectification behaviour as a regular *pn* diode under moderate bias. In contrast, the reverse-biased current abruptly increases approximately 5 orders above a certain threshold voltage (~-4.8 V here), as shown in Fig. 1c. This "hard-knee" rapid change in current signals a typical avalanche breakdown[16, 17] resulting from impact ionization process. Actually, we can exclude a lot of other effects as the original of the "hark knee" breakdown by control experiments (Supplementary Section 1c). It is worth mentioning that the abrupt breakdown phenomenon is very robust. After hundreds of cycles, we can still observe similar avalanche characteristics even operated in breakdown mode (Supplementary Section 2a). To further characterize the nature of reverse breakdown, we have also analysed the transport under different electrostatic doping levels. Avalanche breakdown observed here only occurs in samples with low electrostatic doping, typical Zener breakdown can be observed in highly doped samples (Supplementary Section 2b). Consequently, the swift growth of the reverse-bias current is mainly attributed to avalanche breakdown.

We have also calculated the band profile of our vertical vdW junction to examine the impact ionization process. Fig. 1d shows the vertical band bending at different bias voltages from 0 to -2 V at a fixed $V_{bg}$ of 10 V. The red, green, and blue curves denote the calculated band profiles at biases of 0, -1, and -2 V, respectively. The band structures and absolute alignments of BP and InSe are calculated using hybrid density functional theory (see Methods Section). As shown in the figure, with increasing reverse bias, the band bending and electrical field significantly increase. Even under a 1 V bias, the band tilting deeply exceeds the 1.5 bandgap of BP. This large field is able to highly accelerate carriers, enabling carriers in the narrow bandgap BP to gain enough energy to initiate the impact ionization process. Hence, most of our devices show a very small avalanche breakdown voltage (some even bellow 1 V) (Supplementary Section 2c). In addition, the avalanche characteristics our devices could also show a very small hysteresis window (approximately 0.03 V), which benefits from the perfect interface (Supplementary Section 2d). These unique properties of small hysteresis window and low avalanche breakdown voltage have great benefits for nanophotonics and nanoelectronics. There are two types of devices generally utilize avalanche breakdown: APD and impact ionization MOSFET (IMOS). We therefore experimentally demonstrate the progress in these devices through using the avalanche breakdown.

High efficient light detection elements are important part for mid-IR lidar[11, 12], single photon detection[1] and quantum information[18]. Among the potential candidates, avalanche photodetectors (APDs) in which the photo induced carriers are multiplied by external electric have attracted so much attention due to the high sensitivity. APDs made of our vertical vdW heterostructure present high gain and low noise response for mid-IR light detection. To investigate the photocurrent of the vdW junction, we fabricated the InSe/BP device with an uncovered junction and side electrode, as shown in Fig. 2a inset. The relatively small lateral parasitic resistance of BP does not strongly modify the junction behaviour (Supplementary Section 3). The performances of our APDs have been measured under illumination by a 4 μm laser with in a scanning photo-current microscopy (see Methods Section). Fig. 2a shows the $I_{ds}$-$V_{ds}$ characteristics of a typical APD in dark (grey line and markers) and illuminated (red line) conditions at low temperature. We calculate the multiplication factor, defined[19] by $M = (I_{ph} - I_{dark})/I_{ug}$, where $I_{ph}$ is photocurrent, $I_{dark}$ is dark current and $I_{ug}$ is photocurrent when $M = 1$. Fig. 2a shows a large multiplication factor up to about $3\times10^4$ obtained on breakdown mode at -4.3 V bias. The high multiplication factor greatly benefits to weak light detection. We operated a Fabry-Perot quantum cascade laser below its threshold (in amplified spontaneous emission mode ~50 pW) to demonstrate the weak light detection of our APDs. We estimate that the photon-detection limit is approximately 6,000 photons. We would like to point out the reliable detection limit in terms of photo number is actually limited by the apparatus we used, which is far above the APD's intrinsic limitation. (Supplementary Section 4a).

To qualitatively examine the sensitivity, we measured the current noise density spectra under various reverse biases from 0-4 V, as shown by normalized noise power spectral density, $S(f)/I^2$ in Fig. 2b. This data can be categorized into two parts; the top (bottom) three are the noise power spectral density without (with) avalanche. Interestingly, the noises present a 1/f shape that is distinctively different from the excess noise (white noise) of conventional APDs. The normalized noise power spectral density with avalanche is also much smaller than that without avalanche; detailed analysis indicates that the noise power density of the APDs can be even lower than the theoretical excess noise limit above a certain frequency (Supplementary Section 4b). In semiconductors, the 1/f noise is mainly dominated by the fluctuations of carrier density or mobility. The noise power spectral density $S(f)$ is proportional to $I^2$ with avalanche, which implies that the amplified photocurrent $I$ itself does not drive the fluctuations[20]. This excellent noise property indicates its potential in realizing single photo detection.

Another significant application of avalanche breakdown is IMOS. We engineer IMOS devices based on the vdW heterojunction. The transfer characteristics under different source-drain bias are shown in Fig. 2c. With forward or moderate reverse bias, the transistor behaves as a conventional

MOSFET, where carriers are injected by thermionic emission. Interestingly, the drain current of n-type channel experiences very steep change within a small gate modulation (-2.938 to -2.937 V here) under relatively large reverse bias, representing a small SS of 0.25 mV/dec step even with a 300 nm effective oxide thickness. The observed steep SS is fully consistent with the avalanche phenomenon. Namely, when the channel is reversely biased near breakdown threshold, electrons injected into the channel are apt to acquire enough kinetic energy to initiate impact ionization process. Therefore, superimposing a small gate electric field results in avalanche, amplifying carriers injected in the off-state and leading to steep SS. Obviously, the gate electric field in this IMOS is only used to trigger the avalanche breakdown and does not affect carrier amplification once the carriers are accelerated above avalanche threshold. As a result, an inherently abrupt SS is obtained once the field is sufficient to initiate impact ionization, irrespective of temperature as illustrated in Fig. 2d. The steep SS is obtained over 4 orders of current magnitude, suggesting an impact ionization gain of ~$10^4$, similar as that in APD.

We now move to discuss the mechanism of the impact ionization process of the avalanche breakdown. It should be noted that the avalanche breakdown in our vertical nanoscale vdW junction has 3 unique properties which are fundamentally different from the conventional one.

First, most of our devices show a very small avalanche breakdown voltage, which is visibly superior to conventional avalanche behaviours (from a few to several tens of volts). This is because in our device, the avalanche breakdown occurs in nanoscale. However, conventional avalanche breakdown generally requires a long avalanche region due to the carrier scattering (see below for details).

Second, our devices provide lower avalanche noises which are superior to the theoretical excess noise limit of conventional avalanche. Furthermore, distinctively different from the excess noise (white noise) of conventional APDs, the noise presents a perfect 1/f shape.

Third, in contrast to the conventional avalanche (with a negative temperature coefficient of threshold voltage due to the phonon scattering)[21], our avalanche breakdown shows positive temperature coefficients, and its parasitic resistance limited multiplication factors are almost unchanged in the temperature range from 40-200 K as shown in Fig. 3a and b. The positive temperature coefficients are arising from the Fermi-level broadening and the thermal expansion induced band bending shift (see details below and Supplementary Section 6).

These unique properties suggest a new avalanche mode —— ballistic avalanche. A ballistic avalanche process[22-24] in which the number of ionizing collisions per primary carrier transit is equal to one, had been proposed by Hollenhorst[22] and Jindal[24]. We propose the ballistic avalanche process achieved in the sub-MFP scaled BP channel benefit these impact ionization devices.

For conventional avalanche, a hole gains sufficient energy from an external electric field to impact-ionize atoms in a concatenate manner, generating a large number of free electrons and holes for rapid current multiplication, as shown in Fig. 3c. Because of the scattering of phonons, the impact ionization process occurs randomly, which results in a large excess of noise[3-4]. On the other hand, the ballistic avalanche process is schematically shown in Fig.3d. A hole is accelerated by the electric field to generate an electron-hole pair in a single pass to plane "A"; here, two holes are collected and the electron is injected back to the channel. The electron also initiates impact ionization and generates another electron-hole pair when traveling to plane "B", where two electrons are collected and the hole triggers the loop again. Notably, owing to BP's symmetrical band structure (Supplementary Section 5a), the electron and hole have nearly equal ionization probabilities (both approach unity in the ballistic regime). Without external disturbances, this process is recurrent, contributing a colossal carrier multiplication. In addition, the absence of scattering enables high ionization rates in ballistic avalanche. In this sense, the impact ionization is deterministic once the carrier kinetic energy of BP is higher than the ionization threshold energy. In principle, this deterministic process does not introduce any excess white noise.

The key point for the ballistic avalanche in our device is to accompany avalanche breakdown with a ballistic transport channel. Intuitively, this is possible in the nanoscale BP channel because the strong interlayer coupling[25] of BP contributes to a high mobility in its out-of-plane direction[22-23]. In this scenario, the MFP can be longer than the channel length. To demonstrate this, we calculate the MFPs by using the out-of-plane mobility of BP extracted through the method in Ref 26 (Supplementary Section 5b). Calculation suggests that the MFPs at 200K for holes and electrons are larger than 14 nm and 10 nm, respectively. They are longer than the BP channel length (~10 nm). As a result, the carrier transport is largely ballistic, that is, not obviously affected phonon scattering which contributes to small bias avalanche. Increasing temperature and the thickness of BP, the "hard knee" breakdown disappears due to the channel length is larger than the decreased MFP (Supplementary Section 2e).

We further evidence the ballistic transport nature of nanoscale BP channel by a quantum transport measurement. We sandwich an exfoliated thin BP (~10 nm) between a bottom graphene (few layers) and a top metal electrode to realize ballistic transport in out-of-plane direction of BP as shown in Fig. 4a inset. The detailed fabrication processes are presented in the Methods Section. Different from most of the reported lateral BP devices, carriers transport vertically along the nanoscale thin BP in this structure. We then measure the conductance as a function of back gate voltage $V_{bg}$ and bias voltage $V_{ds}$ at 1.6 K. We observe pronounced quasi-periodic oscillations in 3 devices at low temperature (Fig. 4a and Supplementary Section 7) which indicates the ballistic transport of BP. This oscillation is further verified by measuring the differential conductance ($dI_{ds}/dV_{ds}$) as a function of $V_{ds}$ and of $V_{bg}$, as shown in Fig. 4b. This periodical oscillation of Fabry-Perot type of interferences, has been also observed in carbon nanotube[27, 28] and graphene[29, 30], arise from quantum interference of multiply reflected paths of carrier waves between two partially reflecting electrodes. Both $V_{ds}$ and $V_{bg}$ control the kinetic energy of the carriers and hence tunes their wavelength. This quantum-mechanical wave nature clearly demonstrates that the propagation of carriers in the BP is ballistic — largely free from scattering[27-29]. The ballistic transport verifies that the MFP is larger than channel length. It should be mentioned that the ballistic transport observed here is in a 3D framework, which has been rarely experimentally observed before. Comprehensive understanding of the ballistic transport nature and ballistic avalanche phenomenon call for further deeper theoretical analysis as well as sophisticated transport measurements, which is still an open question at current stage.

In conclusion, we realize avalanche breakdown in a nanoscale vdW heterojunction. The new avalanche phenomenon has small avalanche breakdown voltage (<1 V), negligible hysteresis and excellent repeatability even operated in breakdown mode. Furthermore, we fabricate avalanche photodetectors and impact ionization transistors. The avalanche photodetectors show high photo amplification, sensitive and low noise photodetection for mid-IR light (4 μm wavelength). The impact ionization transistors show steep subthreshold swing (<0.25 mV/dec) and robust stability. Through further quantum transport measurement, we assign the new impact ionization process to a ballistic avalanche process.

**Acknowledgements**

This project was primarily supported by the National Key Basic Research Program of China (2015CB921600 2013CBA01603 and 2018YFA0307200), the National Natural Science Foundation of China (61775092, 61625402, 61574076 and 11374142), the Natural Science Foundation of Jiangsu Province (BK20140017 and BK20150055), the Fundamental Research Funds for the Central Universities, and the Collaborative Innovation Centre of Advanced Microstructures.

**Author Contributions**

X.W. F.M. and A.G. conceived and designed the experiments. A.G. fabricated the devices. A.G. Y.W. J.Z. G.Y and T.C. conducted the transport measurements. A.G. J.L. W.H and D.S conducted the photoresponse measurements. Z.Z. and W.C. performed the DFT calculations. N.W. and X.C. helped grow BP crystals. A.G., X.W and F.M. analysed the data and wrote the manuscript. X.W. F.M. and Y. S. supervised the research. All authors discussed the obtained results.

**Additional information**

Supplementary information is available in the online version of the paper. Reprints and permissions information is available online at www.nature.com/reprints. Correspondence and requests for materials should be addressed to X.W., F.M. and Y. S..

**Competing financial interests**

The authors declare no competing financial interests.

**Data availability**

The data that support the findings of this study are available from the corresponding authors on reasonable request.

**Methods section**

**Device fabrication.** Graphene (InSe) and BP were micromechanically exfoliated on highly p-doped silicon covered by 300 nm silicon dioxide and on polydimethylsiloxane hold by glass slide, respectively. The thickness of graphene (few layers), InSe (~10 nm) and BP flakes (~10 nm) was initially identified by an optical microscope and finally determined by an atomic force microscope (Bruker Multi-Mode 8). The heterostructures of Graphene/BP and InSe/BP were fabricated through the dry transfer method. The whole fabrication process was carried out in a glove box with both oxygen and water concentrations well below one part per ten million (<0.1 p.p.m.). Then, 5 nm Ti/45 nm Au electrodes were fabricated by electron beam lithography patterning, electron-beam evaporation metallization and standard lift-off process.

**Density Functional Theory (DFT) Calculation.** The electronic band structures in out-of-plane direction and absolute alignments of BP and InSe were calculated using density functional theory with a hybrid exchange-correlation functional[31] as implemented in VASP code[32, 33]. Standard values are used for the mixing parameter (0.25) and the range-separation parameter (0.2 A$^{-1}$). (The structures are treated as optimized when none of the residue forces is larger than 0.01 eV/A$^2$. The reciprocal space is sampled with a very fine k-grid[34]. The effective mass of electron (hole) at Γ point is 0.15 $m_0$ (0.28 $m_0$), which is close to the experimental value of 0.128 $m_0$ (0.28 $m_0$)[35]. Notably, owing to BP's symmetrical band structure, the electron and hole have nearly equal ionization probabilities (both approach unity in the ballistic regime).

**Transmission Electron Microscopy.** The TEM-ready samples were prepared using the in situ FIB lift out technique with an FEI Dual Beam FIB/SEM. The samples were capped with sputtered e-C and e-Pt/i-Pt prior to milling. The TEM lamella thickness was ~100 nm. The samples were imaged with an FEI OsirisTF-20 FEG/TEM operated at 200 kV in high-resolution transmission electron microscope mode. The STEM probe had a nominal diameter of 1-2 nm.

**Electrical and optical measurements.** All direct-current electrical characterizations were performed using a Keithley 2636A dual-channel digital source meter. Variable temperature measurements were performed with an Oxford cryostat and a Janis continuous flow optical cryostat system. The differential conductivities were measured using standard lock-in technique (Lock in and current amplifier are Stanford SR830 and Stanford SR560, respectively). The performances of the APDs have been measured under illumination by a 4 μm laser with in a scanning photo-current microscopy. A Fabry-Perot quantum cascade laser (with a centre wavelength of 4.05 μm) was used as a light source. The continuous laser was chopped to specific frequency light which was locked by SR830. The laser beams were focused by a reflection objective. The spot size of the 4 μm wavelength was focused to 8 μm in diameter. The photocurrent was amplified by SR560. The transient response was measured by a 1 GHz bandwidth oscilloscope. The noise was measured by a noise measurement system (PDA NC300L, 100 kHz bandwidth).

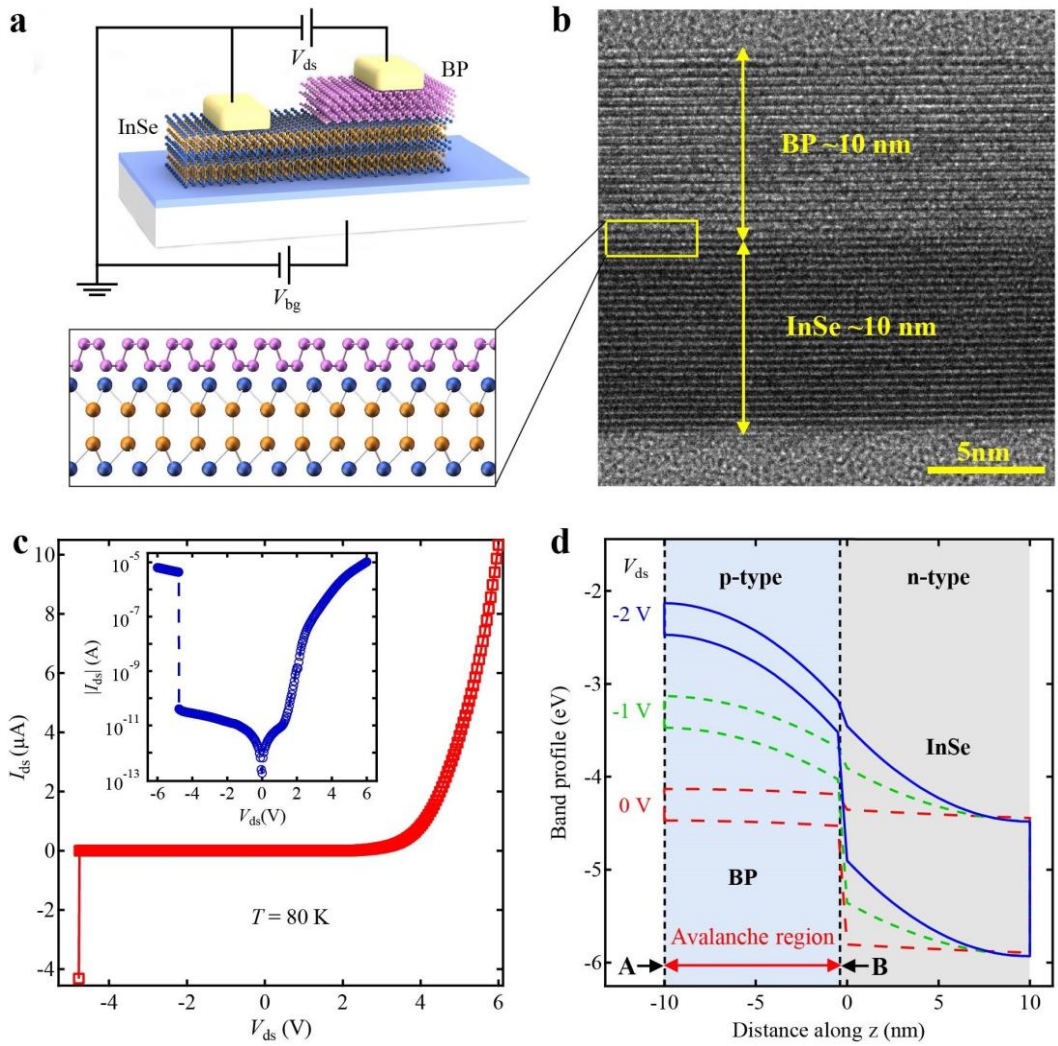

**Fig. 1 Avalanche device of vertical vdW heterojunction. a**, Upper panel: Schematic of the avalanche device of InSe/BP/Metal vertical heterostructures. Bottom panel: Lattice cross-section of InSe/BP interface. **b**. An high-resolution transmission electron microscope image of cross-section of InSe/BP heterostructures. The scale bar is 5 nm. **c**, $I_{ds}$-$V_{ds}$ characteristic of InSe/BP avalanche device in linear scale. Inset: the same $I_{ds}$-$V_{ds}$ curve of main panel in logarithmic scale. **d**, The calculated band profiles at 0, -1 and -2 V bias voltages. Impact ionization occurs in the coloured region (BP). Capitalized letter "A" and "B" denote the surface of BP connected to metal and InSe, respectively.

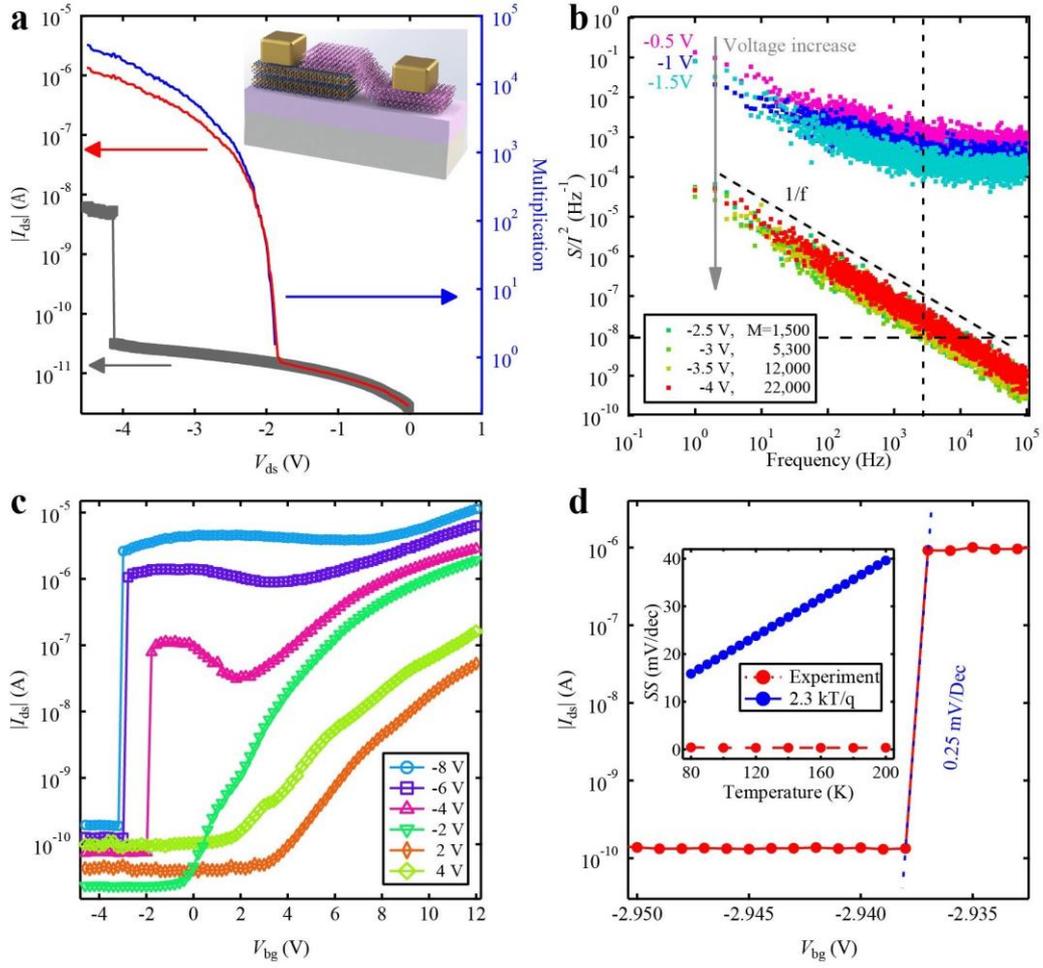

**Fig. 2 Properties of the APDs and IMOS. a**, Low temperature (10 K) photoresponse and multiplication of InSe/BP APDs. Logarithmic scale $I_{ds}$-$V_{ds}$ characteristics (grey line: dark; red line: 4 μm laser illuminated with 30 μW) and corresponding multiplication factor (blue line). The corresponding axes are denoted by the arrows. Inset: Schematic of the APD of InSe/BP heterostructures. **b**, Normalized noise power spectral density (S($f$)/$I^2$) as a function of frequency with different biases under 4-μm laser illuminating at 10 K. The sloping black dash line is a standard 1/f noise guide. The fine dash black transverse line is calculated theoretical noise limit of a conventional avalanche using $S(f)/I^2 = 2eF(M)/I_{ug}$, where $F(M) = 1$. The fine dash black longitudinal line is the boundary of frequency, above which bAPDs have a lower noise level than conventional APDs. **c**, Transfer curves of IMOS at different bias voltages based on BP/InSe heterostructures at 80 K. **d**, The zoom-in view of subthreshold region of transfer curve. the on/off current ratio exceeds $10^4$ and the average subthreshold slope is 0.25 mV/dec. Inset: The average SS at different temperature. The red curve is the experimental value and the blue curve is the fundamental thermionic limitation of SS (2.3 kT/q) at different temperature.

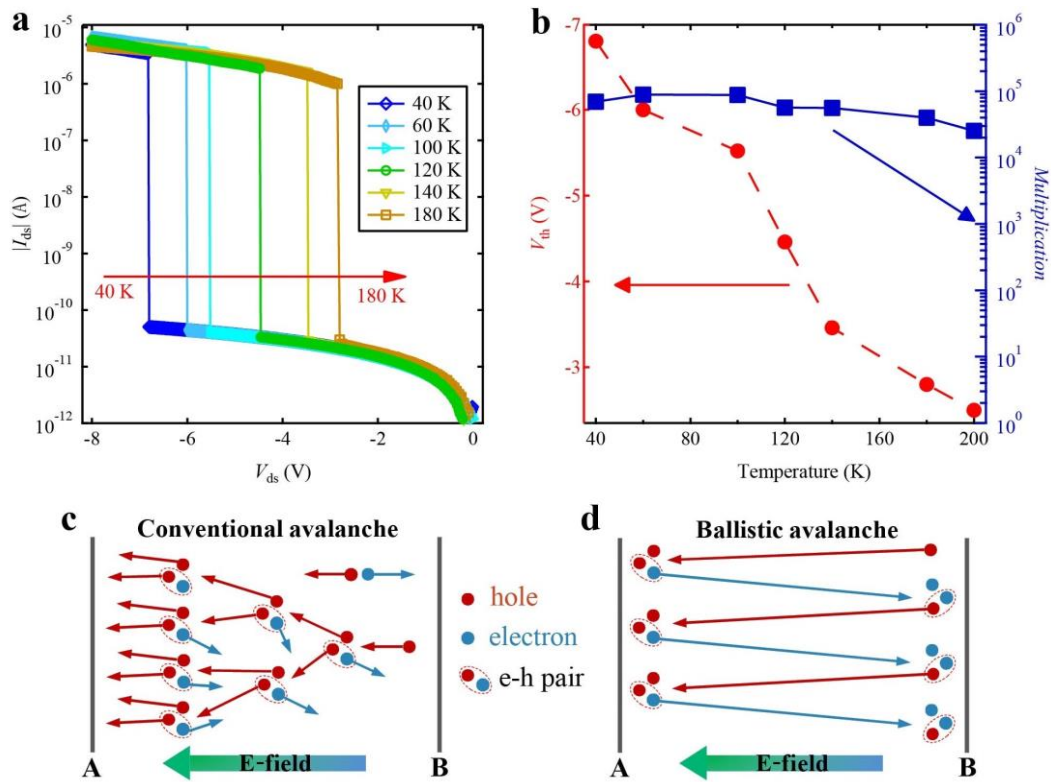

**Fig. 3 Ballistic avalanche mechanism of InSe/BP heterostructure. a**, $I_{ds}$-$V_{ds}$ curves at different temperature (from 40 K to 180 K) in semi-log plot. **b**, Threshold voltages ($V_{th}$) of the avalanche break down and multiplication as functions of temperature. The corresponding axes are denoted by the arrows. **c-d**, Illustration of the conventional (**c**) and ballistic (**d**) avalanche process in the vertical InSe/BP heterostructure devices. Plane "A" and "B" corresponding to top and bottom surface of BP.

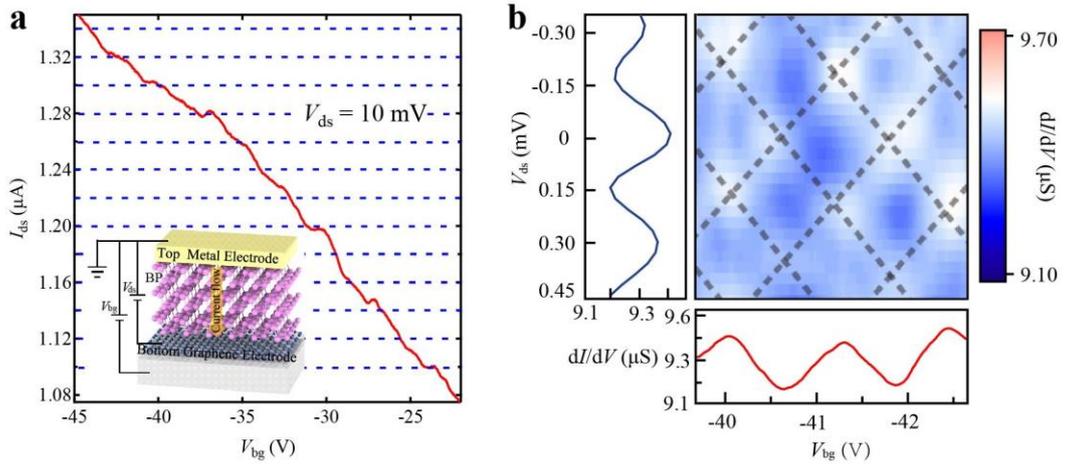

**Fig. 4 Ballistic transport of vertical vdW heterojunction. a,** Conductance $G$ versus back gate voltage $V_{bg}$ at 1.6 K. The conductance shows notable quasi-period oscillation added to a common transfer curve of BP Inset: Schematic of the ballistic transport device of Graphene/BP/Metal vertical heterostructures. The current flow along the vertical direction. **b,** Differential conductance $dI_{ds}/dV_{ds}$ versus $V_{bg}$ and $V_{ds}$ (measured by a lock-in technique) at 1.6 K shows a Fabry-Perot-like interference pattern. Left panel: $dI_{ds}/dV_{ds}$ versus $V_{ds}$ at $V_{bg}$ = -40.6 V extracted from main panel which shows a regular oscillation in a period of 0.3 mV. Bottom panel: $dI_{ds}/dV_{ds}$ versus $V_{bg}$ at $V_{ds}$ = -0.15 mV extracted from main panel.

# Observation of ballistic avalanche phenomena in nanoscale vertical InSe/BP heterostructures

**Anyuan Gao[1], Jiawei Lai[2], Yaojia Wang[1], Zhen Zhu[3], Junwen Zeng[1], Geliang Yu[1], Naizhou Wang[4,5], Wenchao Chen[6], Tianjun Cao[1], Weida Hu[7], Dong Sun[2,8], Xianhui Chen[4,5], Feng Miao[1]\*, Yi Shi[1]\* & Xiaomu Wang[1]\***

---

[1]National Laboratory of Solid State Microstructures, School of Physics, School of Electronic Science and Engineering, Collaborative Innovation Centre of Advanced Microstructures, Nanjing University, Nanjing 210093, China. [2]International Centre for Quantum Materials, School of Physics, Peking University, Beijing 100871, China. [3]Materials Department, University of California, Santa Barbara, CA 93106-5050, U.S.A. [4]Hefei National Laboratory for Physical Science at Microscale and Department of Physics, University of Science and Technology of China, Hefei, Anhui 230026, China. [5]Key Laboratory of Strongly Coupled Quantum Matter Physics, University of Science and Technology of China, Hefei, Anhui 230026, China. [6]College of Information Science and Electronic Engineering, ZJU-UIUC Institute, Zhejiang University, Haining 314400, China. [7]State Key Laboratory of Infrared Physics, Shanghai Institute of Technical Physics, Chinese Academy of Sciences, Shanghai 200083, China. [8]Collaborative Innovation Centre of Quantum Matter, Beijing 100871, China.   \*Corresponding author: E-mail: xiaomu.wang@nju.edu.cn; miao@nju.edu.cn; yshi@nju.edu.cn.

## Supplementary Section 1

**The demonstration of avalanche breakdown**

**a, Electric transport direction of InSe/BP heterostructure**

The transfer curves of InSe and BP are shown in Supplementary Fig. S1. The current of InSe is 1.5 µA ($V_{bg}$ = 10 V) at $V_{ds}$ = 0.1 V. The calculated resistance of InSe is approximately 66 kΩ. The resistance of the junction extracted from Fig. 1c before avalanche breakdown is approximately 245 GΩ, which is approximately $3.7 \times 10^6$ times larger than InSe (66 kΩ) and $3.2 \times 10^3$ times larger than BP (77 MΩ). We therefore conclude that almost all electric voltage is applied to the junction before avalanche breakdown. Once avalanche breakdown happens, the resistance of the junction will diminish quickly, which results in a total resistance (approximately 100 kΩ) similar to the resistance of InSe after avalanche breakdown. Therefore, before avalanche breakdown, the bias voltage mainly applies on the junction between BP and InSe. Once avalanche breakdown occurrence, the external resistances of InSe and BP restrict the increase of current, so we observed the current saturation after avalanche breakdown as shown in Fig. 1c.

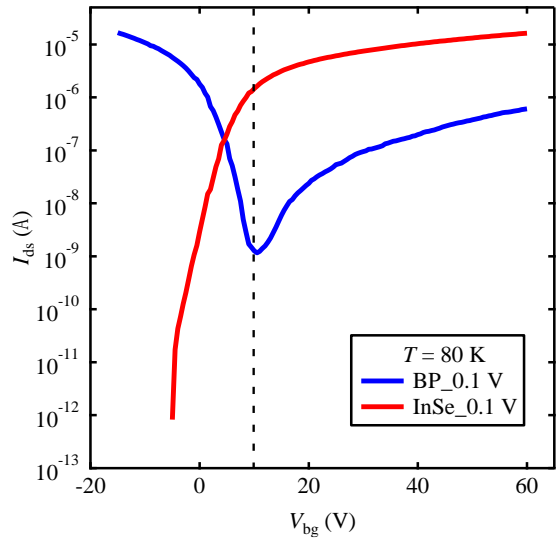

**Fig. S1 The transfer curves of InSe and BP.** Red and blue line are the transfer curve of InSe and BP on 300 nm SiO$_2$ at bias is 0.1 V. The dashed dark line is the back gate voltage where avalanche

breakdown occurs.

**b, Carrier densities of BP and InSe**

The breakdown occurs in Fig. 1c at $V_{bg}$ = 10 V which is the electric neutral point as shown in Supplementary Fig. S1. So the carrier density of electron ($n_e$) and hole ($n_h$) are equal for BP (around charge neutral point). For intrinsic semiconductor, we have[1]

$$n_e n_h = 4\left(\frac{k_0}{2\pi\hbar^2}\right)^3 (m_e^* m_h^*)^{3/2} T^3 \exp(-\frac{E_g}{k_0 T}) \qquad (1)$$

here, $k_0$ and $\hbar$ are Boltzmann constant and Planck constant, $m_e^*$ and $m_h^*$ are effective mass of electron and hole, respectively. $E_g$ is the band gap of BP and $T$ is the temperature. According to equation (1), we get $n_e = n_h \approx 10^8$ cm$^{-3}$.

According to the plane-parallel capacitor model, the surface electron density of InSe ($n_{InSe}$) can be expressed as:

$$n_{InSe} = \frac{\varepsilon_0 \varepsilon_{SiO_2}}{ed}|V_{bg} - V_{th}| \qquad (2)$$

here, $\varepsilon_0$ and $\varepsilon_{SiO_2}$ are vacuum permittivity and relative permittivity of SiO$_2$ (3.9), $e$ is the charge of single electron, $d$ is the thickness of SiO$_2$ (300 nm), $V_{bg}$ and $V_{th}$ are back gate voltage (10 V) and threshold voltage of InSe (2 V here). The electron density of InSe is 6×10$^{11}$ cm$^{-2}$. We estimate the bulk carrier density ~6×10$^{17}$ cm$^{-3}$, by using the surface density divided by the thickness.

**c, Excluding the possibility of other effects**

There are some other effects may result in abrupt current change in 2D materials, such as interfacial trapping effect, phase change, resistive change and polarization change and negative capacitance in ferroelectric material. However, these effects cannot explain the abrupt current change in our devices.

1. The first phenomenon that may induce abrupt current change is trap states. To exclude the key role played by trap states in our heterojunction, we have done high-resolution transmission electron microscope (HRTEM) experiments to characterize the interface between InSe and BP (Fig. 1b). The HRTEM image clearly verifies that the atomic stack is ultra-clean without the presence of any contamination or amorphous oxide even after all the device fabrication processes. In addition, the negligible hysteresis of output curves (Fig. S5) can also support the trap states are minimized in our device. The current level of transfer curve after steep step approaches 10 µA in a few devices, exclude the possibility of the sharp current increase is induced by trap states. Otherwise, the density of trap states should be similar value as carrier density induced by back gate voltage (at the order of $10^{12}$ cm$^{-2}$).

2. Phase change in 2D materials has been also reported to produce abrupt current change[2,3]. There are two ways to induce phase change — charge density wave (CDW) and lattice reconstruction. Generally, the current change induced by phase change should show huge hysteresis due to the change of lattice structure, in contrast the current change in our device shows almost no hysteresis (<0.03 V). Specifically, as far as we know, CDW induced current change is very small and can only be observed in metal. However, the current change in our device shows extremely large "on/off" ratio and can only occur in low-carrier-density BP. Therefore, the abrupt current change observed here cannot be attributed to CDW phase change. Furthermore, structural phase transition induced through electrostatic doping is very difficult. It is only realized in MoTe$_2$ (which have lower phase change energy) with ionic liquid. So it is almost impossible to observe phase change just by SiO$_2$ gate. In addition, structure phase transition does not depend on bias voltage direction which is a contradicting to the observation of abrupt current change only occurs in reverse bias. Accordingly, the abrupt current change observed in our devices is unlikely caused by phase change.

3. Another phenomenon may induce abrupt current change in 2D materials is resistive change[4-6]. It can be divided into volatile type and non-volatile type.   Volatile device[4] can change their resistive state (high resistive state or low resistive state) to initial state once cut off the power supply. Our device obviously presents a 'volatile' behaviour. The key point to realize volatile resistive change is to form a highly movable ion channel (such as using reactive metal contacts). Therefore, volatile resistive change induced current change in our device can also be excluded due to the absence of movable ion channel. Currently, for all the materials we used, only oxidized BP is reported resistive memory device[7]. However, the BP in our device has no interfacial oxide layer (as mentioned above).

4. Abrupt current change induced by polarization change in ferroelectric 2D material[8,9] or Negative Capacitance[10] in ferroelectric dielectrics. However, it does not apply to our device due to no ferroelectric materials is used.

Therefore, the abrupt current increase observed in our experiment is attributed to avalanche breakdown rather than the other effects in 2D materials.

**Supplementary Section 2**

**Novel avalanche breakdown properties of InSe/BP heterostructure**

**a, Repeatability of abrupt breakdown phenomenon**

The InSe/BP heterostructure shows very robust abrupt breakdown phenomenon. After hundreds of cycles, we can still observe similar avalanche characteristics, even operated in breakdown mode. Fig. S2 shows the *I-V* curves before and after hundreds of cycles. No any obvious degeneration is observed after hundreds of cycles except for little change in breakdown voltage and current multiplication.

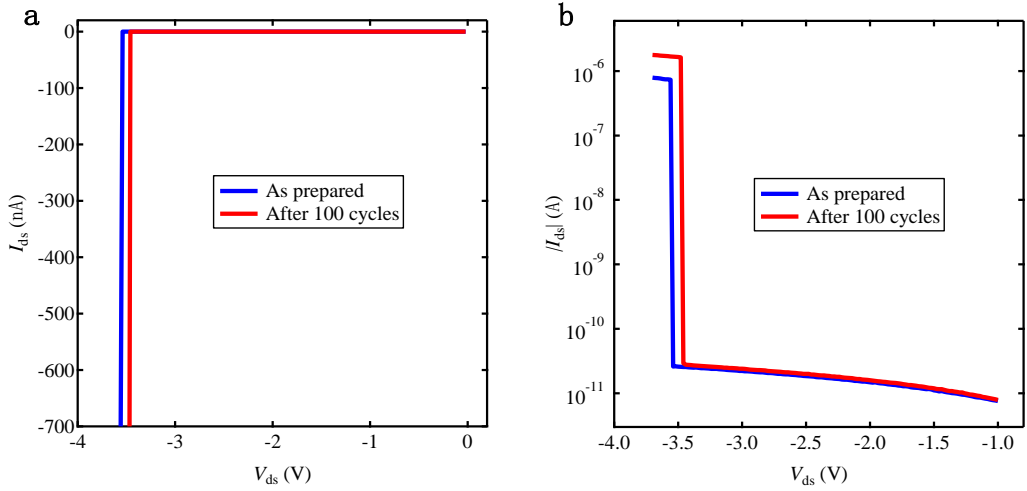

**Fig. S2 Repeatability of abrupt breakdown phenomenon operated in breakdown mode. a**, $I_{ds}$-$V_{ds}$ characteristic of the same InSe/BP device on a linear scale before and after hundreds of cycles. **b**, the same $I_{ds}$-$V_{ds}$ curves of the main panel on a log scale. No degeneration is observed after hundreds of cycles operated in breakdown mode.

**b, Gate tuneable avalanche to Zener breakdown**

Through tuning the carrier density of InSe and BP, we can change of our InSe/BP heterostructure device from avalanche breakdown to Zener breakdown. This phenomenon has also been observed nanowire avalanche device[11]. The gate tunable breakdown property has been shown in Fig. S3, under ultra-low carrier-density condition ($V_{bg}$ = 10 V), device shows avalanche breakdown but Zener breakdown under high carrier-density condition ($V_{bg}$ = 40 V).

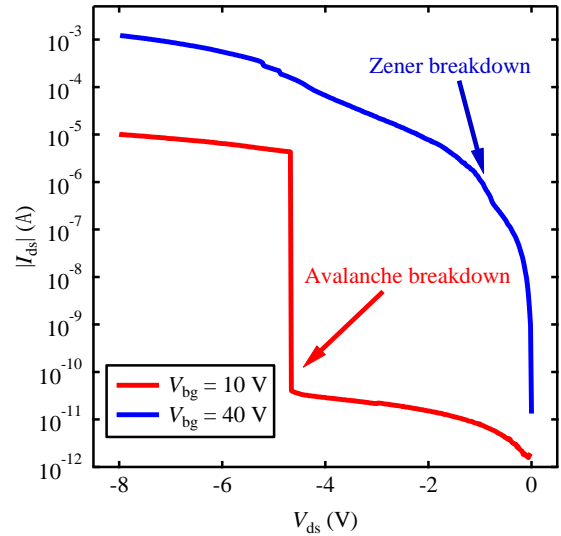

**Fig. S3 Output properties of InSe/BP heterostructures in different doping levels at 80 K.** At a low doping level ($V_{bg}$ = 10 V), the device shows the avalanche breakdown phenomenon. However, at a higher doping level ($V_{bg}$ = 40 V), the device shows Zener breakdown phenomenon.

**c, Ultra-low threshold-voltage breakdown phenomenon**

Our avalanche breakdown also shows ultra-low avalanche breakdown voltage at 240 K as shown in Fig. S4. We find the breakdown voltages in some of our devices (due to the device variation) lower than 1 V which satisfy the requirements of advanced nanoelectronics and integrated photoelectronics (which require the supply voltage lower than 1 V). What's more, this ultra-lower breakdown voltage can't be realized in principle using conventional avalanche breakdown in which to get enough kinetic energy, a very large source-drain bias (usually above few volts) is needed to offset the kinetic energy loss caused by phonon scattering.

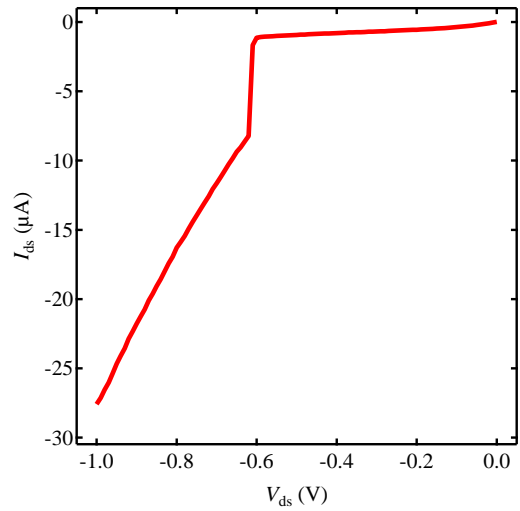

**Fig. S4 *I-V* curve of InSe/BP heterostructures at 240 K.** The threshold voltages of avalanche breakdown are approximately 0.6 V at 240 K, which is much smaller than the conventional avalanche-breakdown voltage.

**d, Negligible hysteresis characteristic of abrupt breakdown phenomenon**

Thanks to the perfect junction interface and high quality lattice structures of InSe and BP, the avalanche characteristics our devices show a very small hysteresis window (approximately 0.03 V), as shown in Fig. S5. These unique properties of ultra-small hysteresis window and low avalanche breakdown voltage have great benefits for nanophotonics and nanoelectronics.

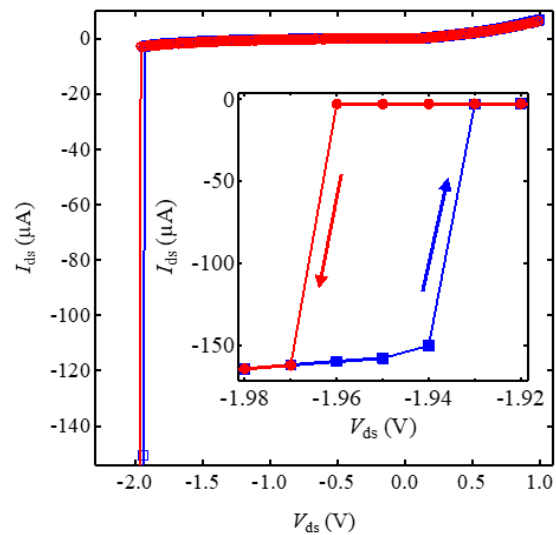

**Fig. S5 Negligible hysteresis characteristic of abrupt breakdown phenomenon.** Hysteresis curve of the avalanche device in another device. Inset: Magnification of the hysteresis curves in the main panel. Sweep directions are marked by arrows.

**e, MFP/Thickness dependent breakdown**

We performed thickness-MFP related experiments to provide further evidence of ballistic avalanche behaviours. First, we investigate the probability of avalanche happing versus sample thickness of BP at a specific temperature of 100 K (Fig. S6a). According to the calculation, the MFP of BP is about 17 nm at 100 K (Fig. S12) which coincide with the sharp decrease of the ballistic-avalanche-happing probability once the BP thicker than 17 nm (Fig. S6a). However, the probability of ballistic avalanche happing is not zero when BP thicker than 17 nm. It is because that for 2D flakes prepared by mechanical exfoliation, the sample quality varies a lot in terms of morphology, mobility and cleanness. Consequently, the MFP largely diverse for different samples.

To further signature the ballistic transport nature, we performed MFP dependent in a fixed device (BP thickness ~10nm). In this structure, we are able to continuously adjust the MFP from longer than the thickness to shorter than the thickness by changing the temperature. The MFP is much longer than the thickness at low temperatures (<100 K) and turns to be shorter than the thickness at ~200 K (Fig. S12). Fig. S6b shows the breakdown curves of the InSe/BP heterostructure as a function of temperature. Obviously, the avalanche breakdown disappears (changes to Zener breakdown) when MFP is shorter than the thickness (>200 K), signalling its ballistic nature. What's more, the avalanche breakdown shows temperature independent property (Fig. 3b) when the channel length lower than (MFP under 200 K). Most of the transport parameters (such as gain and leak current, except threshold voltage) do not change in the ballistic regime.

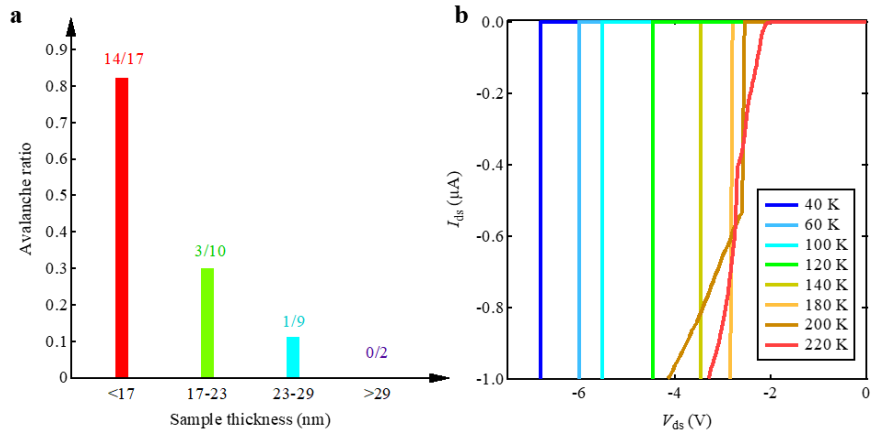

**Fig. S6 Electric breakdown curves vs temperature and BP thickness. a**, The probability of avalanche happening versus the thickness range of BP at 100 K. The left axis is the probability of avalanche happening and bottom axis is thickness range of BP. With the increasing of BP thickness, the probability of avalanche drops quickly. The first and second number of annotations denote the number of samples observed avalanche breakdown and the total number of samples measured, respectively. **b,** The I-V curves of InSe/BP heterostructure at different temperatures. As temperature increases, the current breakdown change from avalanche breakdown to Zener breakdown.

## Supplementary Section 3

**Avalanche breakdown comparison for horizontal and vertical InSe/BP heterostructure**

Horizontal and vertical InSe/BP heterostructure are shown in Fig. S7a inset. BP is covered on InSe denoted by red arrow. The structure of electrodes 1-2 denote vertical heterostructure. The electrodes 1-3 denote horizontal heterostructure which is benefit to determine the photocurrent of the heterostructure. The relatively small lateral parasitic resistance of BP does not strongly modify the junction behaviour. So, the electrodes with covered and uncovered junctions both show similar avalanche breakdown phenomena in spite of the small difference in avalanche-breakdown voltages.

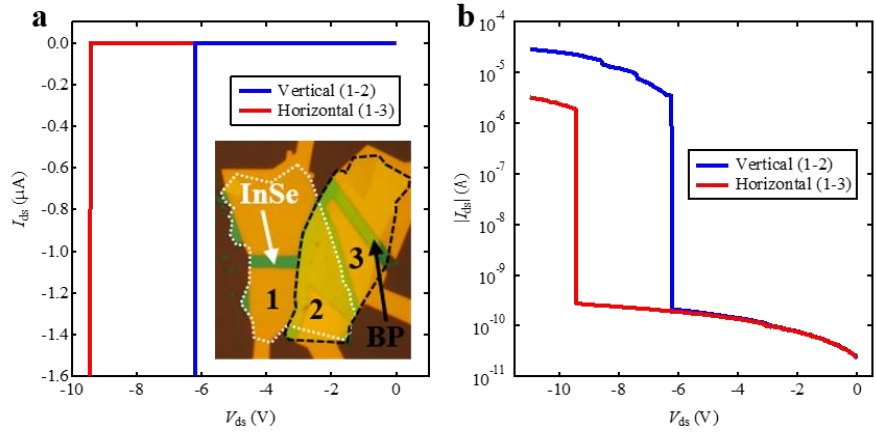

**Fig. S7. Avalanche breakdown phenomena of InSe/BP device with covered and uncovered junction at $V_{bg}$ = 10 V. a,** Inset: Microscope image of the InSe/BP device. The electrodes are numbered 1-3. The sample of bottom InSe and top BP are indicated by white and black dashed line, respectively. Main panel: *I-V* curves of the vertical (electrodes 1-2) and horizontal (electrodes 1-3) junction in blue and red colour lines, respectively. **b,** Output properties of the junction corresponding to the left figure on a semi-logarithmic scale.

## Supplementary Section 4

**Photodetection of the avalanche breakdown**

**a, Estimation of detection limit of InSe/BP APDs.**

We operate a Fabry-Perot quantum cascade laser below its threshold (in amplified spontaneous emission mode) to demonstrate the ultra-weak light detection of the bAPD. To fulfil the rise time of the bAPD (150 μs, Fig. S8), laser pulse was chopped to a frequency of 6.7 kHz. The power of the 4 μm laser was measured by standard liquid nitrogen-cooled HgCdTe photodetector (FTIR-16-2.00, Infrared Association, Inc.). The laser is then attenuated to the detection limit of this HgCdTe photodetector. The total incident power is estimated about 820 pW at 6.7 kHz bandwidth according to its NEP (10 pW Hz$^{-1/2}$); or equally 2.4×10$^6$ photons per pulse for 4 μm laser illumination. The effective detection area of the bAPD used is approximately 2×2 μm$^2$, and the laser is focused to a spot size of 8×8 μm$^2$. So, we get the power illuminating on our device is ~50 pW. According to the absorption spectrum of BP (approximately 10 nm)[12], only 4% of the incident light is absorbed at 4 μm wavelength. Therefore, 0.25% of total incident light is absorbed by the bAPD. Taking all these factors into consideration, we conclude that our bAPDs can produce measurable photocurrent for ~6,000 incident photons (2.4×10$^6$ ×0.25%).

In reality, this is a conservative estimate, i.e. the real minimum detachable photon number is smaller, because the response speed is limited by external circuits rather than photo-generated carriers' lifetime and the light powers are not weak enough to the detection limit of our APDs (the photocurrent is still well above the noise).

We also use the experimental NEP to calculate the theoretical detection limit of our bAPDs. We extract the noise (570 pA) for M=1500 from Fig. S9 at 6.7 kHz. The true detection limit is thus 0.38 pA (at 6.7 kHz bandwidth) when signal to noise ratio equals to 1, corresponding to a minimum detectable limitation of 354 photons.

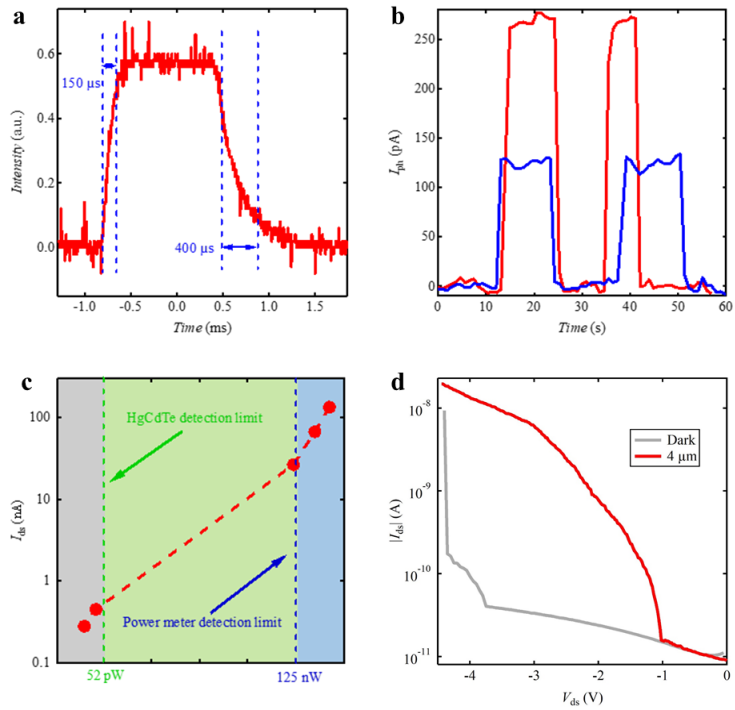

**Fig. S8 Ultra-weak light response of 4 μm wavelength at 10 K. a,** The rise/fall time are defined as from 10/90% to 90/10% of the stable photocurrent after turning the laser on/off. The rise time of 150 μs and the fall time of 400 μs were obtained with a 411 Hz chopping frequency. The relatively slow speed is attributed to a large parasitic resistance. Improving lateral parasitic resistance is expected to decrease the response time of the bAPDs. **b,** The photocurrent $I_{ph}$ as function of time under modulated light on and off. The blue and red curves are the photoresponse under different light power illuminating (50 and 30 pW for red and blue lines, respectively). The powers are both estimated according to the detection limit of HgCdTe photodetector at 77 K. These values only have relative meaning. **c,** The relationship between photocurrent and laser power. The dashed green and blue lines are the detection limit of HgCdTe (liquid nitrogen cooling) and the power meter, respectively. The left two dot below 52 pW corresponding to **b**. **d,** The *I-V* curves of InSe/BP ballistic avalanche photodetector with ultra-low power illuminating which lower than the detection limit of HgCdTe photodetector at 77 K.

**b, The ultra-low current noise of InSe/BP APDs**

To qualitatively examine the sensitivity, we measured the current noise density spectra under various reverse biases from 0-4 V (Fig. S9). The current noise levels show similar intensities before avalanche breakdown but grow quickly with increasing amplification factor. More interestingly, the noises present a perfect 1/f shape that is distinctively different from the excess noise (white noise) of conventional APDs due to the ballistic avalanche's deterministic nature. For comparison, we calculate the excess noise of conventional avalanche photodetectors using the following formula[13]:

$$S(f) = 2eI_{ug}M^2 F(M) \qquad (3)$$

$$F(M) = M\left[1 - \left(\frac{1-k}{k}\right)\left(\frac{M-1}{M}\right)^2\right] \qquad (4)$$

where $e$ is the charge of a single electron, $I_{ug}$ is the photocurrent when $M = 1$, $M$ is the current multiplication factor, $F(M)$ is the excess noise factor and $k$ is the ionization coefficient defined as $\alpha_h/\alpha_e$. The theoretical excess noise limits (defined as $F(M) = 1$), denoted by dashed lines, have been plotted for different values of $M$ in Fig. S9. The same multiplication factor is drawn in the same colour. Remarkably, the noise power density of the bAPDs can be even lower than the theoretical excess noise limit above a certain frequency.

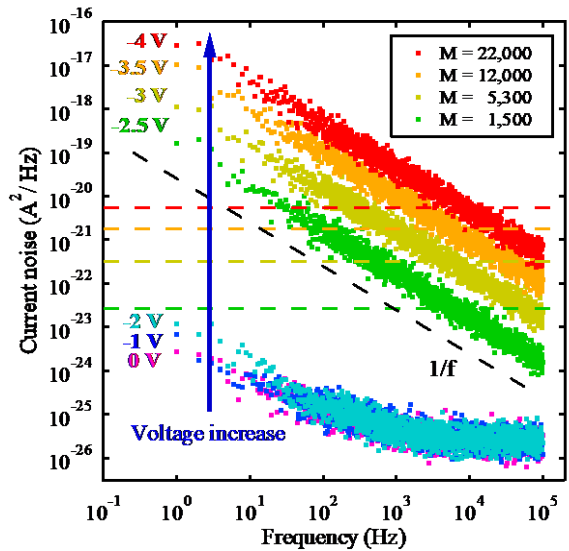

**Fig. S9 Noise power spectral density (*S*) of APDs with different biases under 4-μm laser**

**illuminating at 10 K.** The black dash line is a standard 1/f noise guide. The dashed colour lines are the theoretical noise limits of conventional APDs at $M = 1$ (with same colour as multiplication).

## Supplementary Section 5

**The high mobility of BP in out-of-plane direction**

**a, The out-of-plane effective mass of BP**

The experimental bulk value of the effective mass ($m^*$) of BP in the out-of-plane direction are 0.28 $m_0$ and 0.128 $m_0$ for hole and electron, respectively[14-16]. We calculate the band structure of 10 nm BP along vertical direction (Supplementary Fig. S10), the $m^*$ of hole (electron) at Γ point is 0.28 $m_0$ (0.15 $m_0$), which are close to the experiment value.

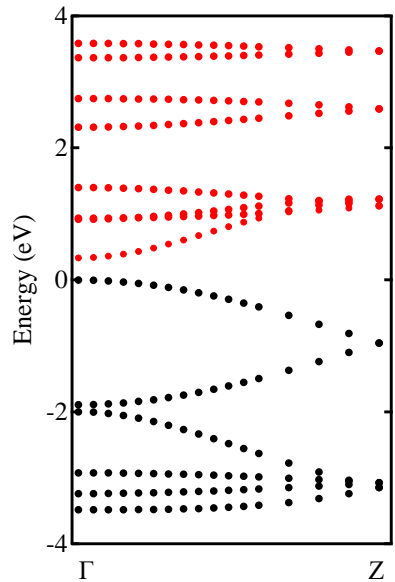

**Fig. S10 The calculated band structure of BP in z direction.** The Fermi energy is set as 0 eV. The effective mass of electron (hole) at Γ point is 0.15 $m_0$ (0.28 $m_0$), which is close to the experimental value of 0.128 $m_0$ (0.28 $m_0$).

**b, Carrier (electron and hole) mobility and Mean Free Paths (MFPs) of BP**

We fabricated graphene/BP/metal device to measure the carrier density of BP in vertical direction as shown in Supplementary Fig. S11a inset. The thickness of BP is ~16 nm. The overlapping area of metal and graphene is ~1 μm². The out-of-plane mobility of BP can be fitted by $\mu = \frac{J \cdot L}{neV_{ds}}$, where $J$ is the current density, $n$ is carrier density, $e$ is the charge of an electron and $L$ is the thickness of BP[17]. We use equation (1) to calculate the room temperature intrinsic carrier density of BP (~5×10²¹ m⁻³). By using experimentally extracted $J$ = 600 μA/μm², $L$ = 16 nm, $n$ = 5×10²¹ m⁻³ and $V_{ds}$ = 0.42 V (Fig. S11a), we extract the room temperature mobility of BP in out-of-plane direction as ~320 cm² V⁻¹ s⁻¹ which is comparable to the in-plane-mobility (Fig. S11b). If we conservatively assume the scattering lifetime is dominated by phonon (50-300 K) and scales as $T^{-0.9}$, where $T$ is the temperature, the mobility at different temperature can be expressed as $\mu(T) = 5.5 \times 10^4\ T^{-0.9}$ (cm² V⁻¹ s⁻¹) (Fig. S11b).

At high-temperature region, the mean scattering is phonon scattering. The mean free path dominated by phonon scattering can be expressed as the following equation[18]:

$$l = \frac{3\sqrt{2\pi m^* k_0 T}}{4e}\mu \qquad (5)$$

Here, $l$, $\mu$ and $m^*$ are the mean free path, mobility and effective mass of electron or hole. $k_0$ is the Boltzmann constant and $T$ is the temperature. The effective masses of electrons and holes along the z-direction are 0.128 and 0.280 $m_0$ (ref [14]), respectively, where $m_0$ is the free-electron mass. We use the experimental mobility of room temperature to calculate the mean free path. With temperature bellow 200 K, the MFPs are larger than 10 nm both for electron and hole. But with temperature higher than 200 K, only the MFP of hole is larger than 10 nm (Fig. S11c).

We also use thermal velocity to estimate the MFP at low temperatures. The thermal velocity ($v$) is given by $v = \sqrt{3kT/m^*}$, where $k$ is Boltzmann constant, $T$ = 1.6 K here. So the thermal velocities for holes and electrons are 1.6×10⁶ cm s⁻¹ and 2.3×10⁶ cm s⁻¹, respectively. The MFPs ($l_{MFP}$) are about 110 and 160 nm ($l_{\text{MFP}} = v\tau$) for holes and electrons, which are considerably larger than the channel length of BP (~10 nm).

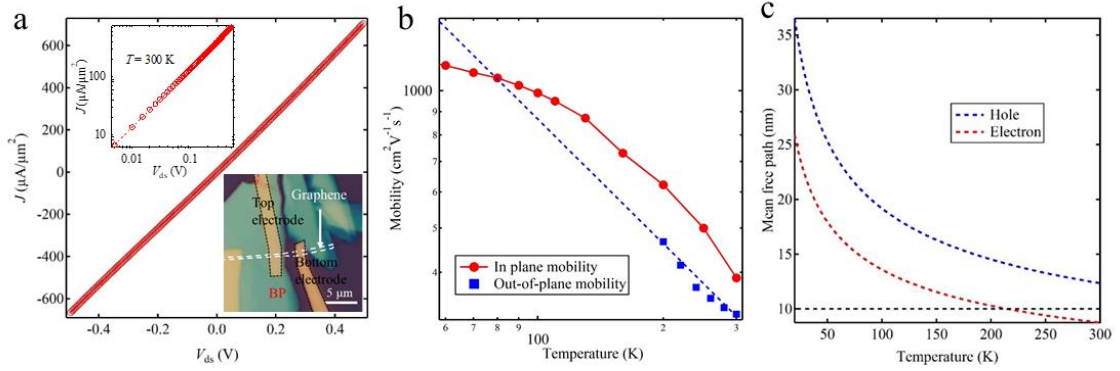

**Fig. S11 *IV* curve, carrier mobility and mean free path of BP in out-of-plane direction. a**, *IV* curve of graphene/BP/metal heterostructure at room temperature. Bottom inset: Optical microscope image of graphene/BP/metal heterostructure. The thickness of BP is ~16 nm. The overlapping area of metal and graphene is ~1 μm². Top inset: *IV* curve of graphene/BP/metal heterostructure at 300 K in log scale. Main panel: *IV* curve of graphene/BP/metal heterostructure in linear scale. **b**, Mobility of BP as a function of temperature. In plane mobility of BP in InSe/BP heterostructure is denoted by red curve. The out-of-plane mobility of BP is depicted by blue curve. It is fitted by $\mu \sim T^{-0.9}$ and the room temperature mobility is extracted from $\mu = \frac{J \cdot L}{neV_{ds}}$. **c**, Carrier mean free path as function of temperature. The red and blue dashed lines are the MFPs of electron and hole, respectively.

## Supplementary Section 6

**Anomalous positive temperature coefficient of threshold voltage of ballistic avalanche breakdown**

The ballistic avalanche presents an abnormal temperature dependence of the threshold voltage. In contrast to the conventional avalanche (with a negative temperature coefficient), our ballistic avalanche phenomenon shows positive temperature coefficient. The positive temperature coefficient has been observed in InGaAs system also with low band-gap and small carrier effective mass[19,20]. It is attributed to a phonon-assisted avalanche process. However, our device has a different physical origin.

In the phonon-assisted avalanche, phonons participate into the impact ionization process by carrying away a minimum momentum of ionizing particles around energy valley centre. The number of final ionizing states increases with increasing temperature for phonon-assisted avalanche, resulting in a decreased threshold voltage. In certain materials (which are with small bandgap and low effective mass), phonon-assisted avalanche turns pronounced and even dominate over conventional avalanche. As a result, positive temperature coefficient of avalanche threshold voltage is observed[19,20]. Obviously, phonon-assisted avalanche requires phonon to strongly involve in carrier transport. There are several evidences support that in our device, electrons almost do not collide with phonons in the ballistic avalanche regime: First, the ultrathin channel thickness is shorter than the MFP. Second, the mobility presents a typical phonon limited behaviour, and phonon scattering is largely suppressed at low temperatures. Third, with phonon scattering, the ultrathin channel length hardly generates so large current multiplication. Therefore, the positive temperature coefficient in our devices can't be explained by phonon-assisted avalanche.

We found our positive temperature coefficient is due to two reasons. First, as temperature increases, the Fermi-Dirac distribution expands to higher energy, number of carriers with higher energy will increase. Hence, with smaller bias, the carriers are also able to get enough energy to initiate the avalanche breakdown. This behaviour indicates that the carriers undergo similar acceleration and impact ionization processes at different temperatures without being significantly affected by phonons. Second, the thermal expansion induced band bending shift. Briefly, we find the built-in potential of the BP/InSe junction increases with temperature, due to InSe's anomalous thermal expansion[8] (which is negative thermal expansion coefficient in low temperature) and BP's negative Poisson's ratio[9] (which results in a thermal induced band gap opening). For simplicity, we mainly consider thermal expansion contribution in calculating temperature dependent band structures. After that, we derive the band shift (as a function of temperature) by

$$\Delta E_k(T) = -\frac{1}{\chi}\frac{\partial E_k}{\partial P}\int_0^T \beta(T)dT \qquad (6)$$

where $\beta$ and $\chi$ are volumetric thermal expansion coefficient and compressibility, respectively. The main results are demonstrated in Fig. S12. It should be noted that the calculation largely reproduces materials' thermal properties, including BP's unique temperature-induced band gap opening[21]. In addition, the extracted band bending increases with increasing temperature, suggesting smaller external voltage is required to initiate the impact ionization process. This is because compared with BP, InSe presents a much smaller thermal expansion coefficient (even be negative for 50-100K region)[22]. Its conduction band edge only slowly changes with respect to the valence band edge of BP, raising the built-in potential.

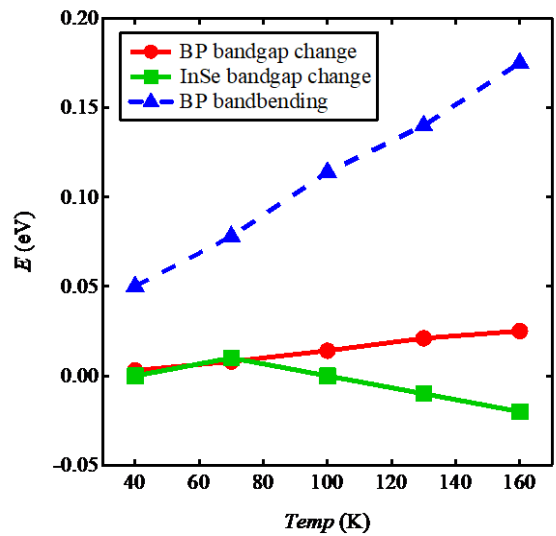

**Fig. S12 The relationships of bandgap changes and band bending with temperature.** The green and red solid lines are the bandgap changes of InSe and BP relative to the bandgap at 0 K. The blue dashed line is the band bending of BP. The bandgap of BP (InSe) increase (decrease) with the increase

of temperature. The band bending of BP increase with temperature increasing which resulting in positive temperature coefficient.

## Supplementary Section 7

**The quasi-periodic current oscillation in another 2 devices**

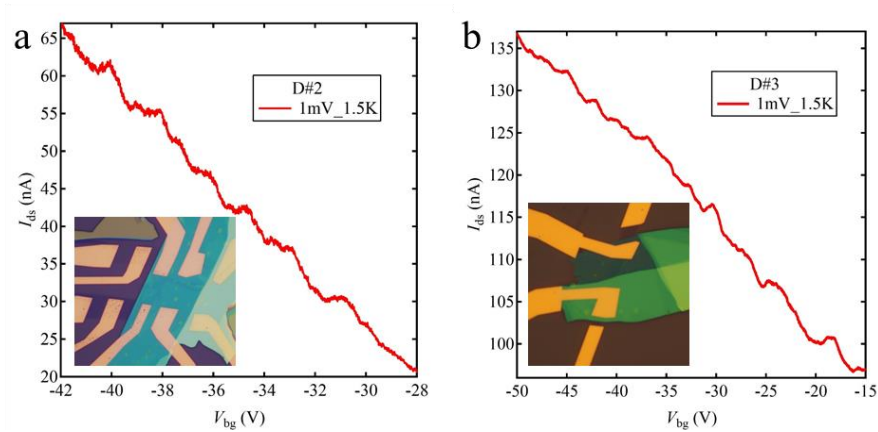

**Fig. S13 Quasi-periodic current oscillation of Device #2 and #3 at 1.5 K.**

## Supplementary Section 8

**Photoresponse in reverse and forward bias voltages**

To further confirm the photoresponse in reverse bias voltage caused by avalanche multiplication, we measured the photoresponse in one of our device with low-power-laser illumination. Fig. S14 shows the $I_{ds}$-$V_{ds}$ curves with (red curve) and without (grey curve) 1550-nm-laser illumination. In dark condition, this device shows notable avalanche behaviour. With a low power light illumination, a notable response is observed in reverse bias while without any response is observed in forward bias. This means that no any notable photocurrents generate without avalanche occurrence. Therefore, the photocurrents observed in reversed bias voltages come from avalanche photo amplification.

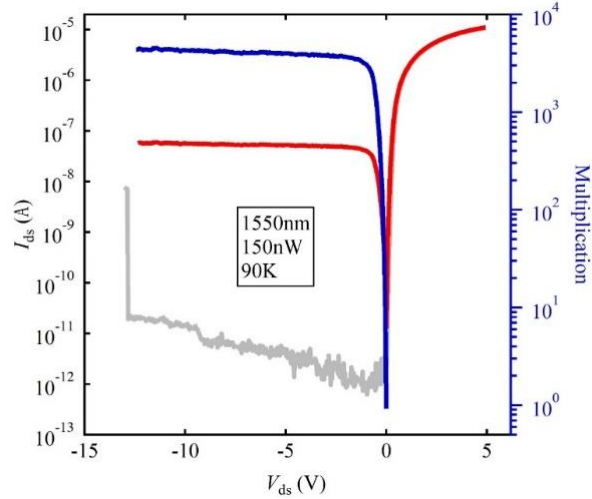

**Fig. S14 Photoresponse in reverse and forward bias voltages of APDs at 90 K.** $I_{ds}$-$V_{ds}$ curves with and without light illumination are denoted by red and grey curves, respectively. The blue curve is the multiplication corresponding to right axis. The photocurrents are measured with a 1550-nm laser at 150 nW.

**Supplementary Section 9**

**The photoresponse of InSe/BP heterostructure phototransistor**

In order to directly distinguish the origin of the multiplication, we measure the photoresponse of InSe/BP heterostructure with and without avalanche in the same device. As shown in the main text, the photo-gain is very large in the ballistic avalanche regime. Decreasing the MFP by increasing temperature, the ballistic avalanche disappears in a same device, permitting us to measure the response of the same device without avalanche. Fig. S15 is the photoresponse without avalanche under 1.4 μW laser illumination. This photocurrent increases with the increase of electric filed, but with much lower EQE (<0.1) which is much lower than the EQE of avalanche breakdown (~80). The distinctively different photoresponse in a same device with and without avalanche verify that the photomultiplication observed here is due to avalanche multiplication.

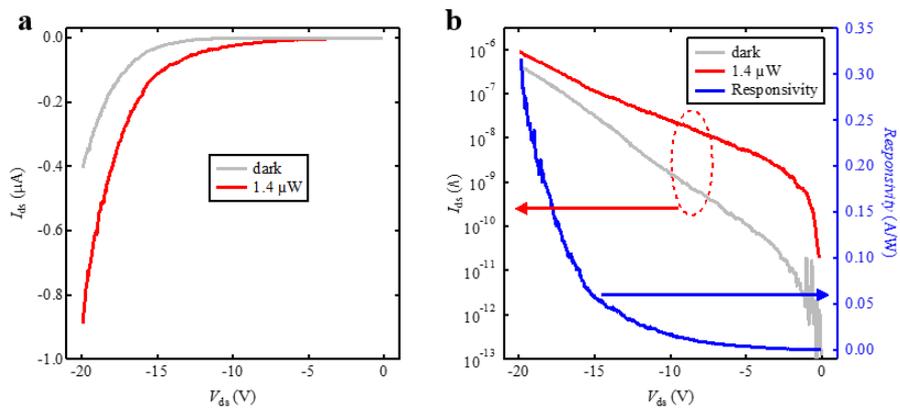

**Fig. S15 Photo-response properties of InSe/BP heterostructures without avalanche under 4 μm laser illumination at 80 K. a,** *I-V* curves of the heterostructures on a linear scale. Dark and illuminated conditions are denoted by grey and red, respectively. **b,** photocurrent (left axis) corresponding to left figure on a logarithmic scale and responsivity (right axis).